\documentclass[usenatbib]{mn2e}

\usepackage[english]{babel}
\usepackage[latin1]{inputenc}
\usepackage[T1]{fontenc}

\usepackage{amsmath}
\usepackage{bbold}
\usepackage{graphicx}
\usepackage{amsfonts}
\usepackage{amssymb}
\usepackage{hyperref}
\usepackage{latexsym}

\newcommand{\be}{\begin{equation}}
\newcommand{\ee}{\end{equation}}
\newcommand{\bea}{\begin{eqnarray}}
\newcommand{\eea}{\end{eqnarray}}

\begin{document}


\title[Scaling of the 1-halo terms with bias]{Scaling of the 1-halo terms with bias}

\author[L.~R.~Abramo \& Ir\`ene Balm\`es \& Fabien Lacasa \& Marcos Lima]
{L.~Raul Abramo$^1$\thanks{abramo@if.usp.br}, Ir\`ene Balm\`es$^1$, Fabien Lacasa$^{2}$, 
Marcos Lima$^1$
\\
$^1$ Departamento de F\'{\i}sica Matem\'atica,
Instituto de F\'{\i}sica, Universidade de S\~{a}o Paulo, 
CP 66318, CEP 05314-970 S\~{a}o Paulo, Brazil
\\
$^2$ ICTP South American Institute for Fundamental Research, 
and Instituto de F\'{\i}sica Te\'orica
\\ 
\hskip 0.2cm Universidade Estadual Paulista (UNESP), 
Rua Dr. Bento Teobaldo Ferraz 271, 01140-070 S\~ao Paulo, 
SP Brazil}

\maketitle

\begin{abstract}
In the Halo Model, galaxies are hosted by dark 
matter halos, while the halos themselves are biased tracers 
of the underlying matter distribution. Measurements of  
galaxy correlation functions include contributions both from 
galaxies in different halos, and from galaxies in the same 
halo (the so-called 1-halo terms).
We show that, for highly biased tracers, 
the 1-halo term of the power spectrum obeys a steep scaling 
relation in terms of bias.
We also show that the 1-halo term of the trispectrum has a steep
scaling with bias.
The steepness of these scaling relations is such that
the 1-halo terms can become key contributions to the
$n$-point correlation functions, even at large scales.
We interpret these results through analytical arguments and 
semi-analytical calculations in terms of the statistical properties 
of halos.
\end{abstract}


\section{Introduction}


Galaxy surveys \citep{York:2000gk,cole_2df_2005,Abbott:2005bi,scoville_cosmic_2007,PAN-STARRS,BOSS,2011MNRAS.415.2876B,Andersonetal2012,Andersonetal2014} 
are not just tools for constraining cosmological parameters: 
they reveal the 3-dimensional spatial web of visible structures, the time evolution of
these structures, and the history of galaxy evolution.
Next-generation surveys are aimed at answering a 
variety of open astrophysical and cosmological questions, by collecting 
vast amounts of data at low, intermediate, and high redshifts 
\citep{BigBOSS,Benitez:2008fs,Benitez:2014,PFS,eBOSS,DESI,LSST:2009pq}.

For the study of how gravity and the Universe's background expansion affect the growth of structure,
the standard statistical tools are the $n$-point correlation functions and their 
Fourier transforms, the polyspectra. If the distribution of matter were perfectly Gaussian,
the two-point correlation function (2pCF) or, equivalently, the matter power spectrum, 
would contain all the statistical information.
However, the primordial fluctuations are believed to be very nearly, but not perfectly, Gaussian. 
Information about the processes that generated these primordial perturbations is encoded 
in higher order moments, e.g. the bispectrum 
\citep{Maldacena2003}. Moreover, when subject to non-linear time evolution, 
even a perfectly Gaussian initial field develops nontrivial 
higher-order moments (skewness, kurtosis, etc.) whereto statistical 
information propagates. At late stage of non-linear gravitational evolution, information 
even leaks out of the hierarchy of moments, as the density field becomes 
approximately lognormal \citep{Carron2011,Carron2012a,Carron2012b,Carron2014}.

Adding to these complications is the fact that we do not directly observe the   
total matter distribution, but only its visible component --- which accounts to 
$\sim 20$\% of the total matter \citep[see, e.g.][]{Planck2015-cosmo}
and is affected by non-gravitational effects
such as the physics and feedback of baryons, radiation pressure, etc.
Hence, the 2pCF of the distribution of visible matter cannot possibly tell the 
full story 
and we must treat galaxies, quasars, 
Ly-$\alpha$ systems, etc., as unfaithful 
(and biased) tracers of the underlying dark matter (DM) distribution.

The relation between tracers of large-scale structure (LSS) and the DM distribution is
partially provided by the Halo Model \citep{CooraySheth}. 
The DM halos --- and not the DM particles --- then become the fundamental objects.
In particular, the correlation functions of the DM halos are related 
to the correlation functions of the DM particles by the halo abundance, bias and profile, 
such that more massive halos are less abundant, are more highly biased and have less concentrated profiles.

The statistics of how galaxies populate DM 
halos is provided by the so-called Halo Occupation Distribution (HOD) 
\citep{MartinezBook,Berlind2002,Kravstov2004,Zheng2005}, which specify how many observable galaxies inhabit a halo as a function of its mass. 
HOD parameters can be calibrated by measurements of specific
abundance ratios, and by fitting the observed correlation functions.

The relationship between the statistics of the halo density field and that of galaxies is often non-trivial.
For example, the $n$-point statistics of galaxies get contributions from 
the $N$-halo term (when each galaxy occupies a different halo), 
from the $(N-1)$-halo term (when two galaxies occupy the same halo, 
and the others lie in different halos), etc., all the way down to 
the 1-halo term (when all the $N$ galaxies occupy the same halo). 
Furthermore, there are contributions to the $N$-point statistics
from $N-1$ types of shot-noise terms: 
e.g., the $(N-1)$-halo -- $(N-1)$-galaxy term (when the correlation function hits 
twice a galaxy in a given halo, and then only once galaxies in 
different halos) down to the 1-halo -- 1-galaxy term 
(when the correlation function hits $N$ times the same galaxy). 
Hence, when we measure 
the $N$-point function of galaxies, we are in fact measuring an 
ad-mixture of all the $N'$-point functions 
of halos ($N'= 1, 2, \ldots, N$).

The galaxy 2pCF includes both a 2-halo term (related to galaxies in two distinct halos) and a 1-halo term (accounting for galaxies within the same halo). 
Clearly, the 2-halo term is most important on large scales, within the linear regime, and bears the imprint of the large-scale matter distribution, 
whereas the 1-halo term dominates on small, nonlinear scales, 
and reflects the matter distribution inside halos (the halo density profile). 
Even though these two regimes are connected by time-evolution, we may 
describe the galaxy power spectrum as the superposition of 
two independent scaling laws: that of the linear DM power spectrum, 
which dominates on large scales, and that coming from the halo profiles, 
which dominates on small scales ($k \gtrsim 1 \, h$ Mpc$^{-1}$). 

In the large-scale limit the 1-halo term contributes 
a constant to the galaxy power spectrum. This constant may be 
poorly known, since galaxy surveys designed for cosmological studies are 
often insufficiently complete (or have poor redshift accuracy, in 
case of photometric redshifts) to determine precisely the HOD parameters 
for a type of galaxy, and for the survey's mean 
redshift. This constant represents a noise that must 
be subtracted from the power spectrum --- just as it happens 
with shot noise, which has a completely different origin but is 
also a constant that must be subtracted. 
This feature of the 1-halo term on large scales also comes into play 
in higher-order statistics: the bispectrum gets a constant contribution 
from the 1-halo term of the 3-point function; the trispectrum gets 
a constant contribution from the 1-halo term of the 4-point function; 
and so on. 

In this paper we study the behavior of the 1-halo terms as a function 
of the galaxy bias, 
where these quantities are connected via their mutual dependence 
on HOD parameters.
The same goes for the other observable 
quantities, such as the mean number density of galaxies, the
2-halo term of the power spectrum, etc. 
Hence, the HOD provides an internal (but unobservable) 
parameter space that we can use to vary the observable quantities.
In particular, we use the galaxy bias to parametrize the 1-halo terms both 
because it is more readily available in observations, and also 
because we are interested in identifying universal behaviors, 
regardless of the details of the HODs.

We show that, for highly biased tracers ($b_g \gtrsim 3$), 
the 1-halo term of the 2pCF obeys a scaling relation 
in terms of bias, growing as $P^{1h} \sim b_g^{4-5}$, which is 
much faster than the scaling of the 2-halo term 
($P^{2h} = b_g^2 P_m$, where $P_m$ is the matter power spectrum). 
For highly biased galaxies, the effective shot noise contributed by the 1-halo term 
can become at least comparable to the Poisson shot noise, significantly 
lowering the signal-to-noise ratio for measurements of the power 
spectrum, baryon acoustic oscillations, etc. 
In some cases the 1-halo term can even surpass by a large factor the shot noise, as is the case, 
e.g., for the angular power spectrum of the cosmic infrared background on the angular 
scales probed by Herschel \citep{Thacker}.
Furthermore, we show that the 1-halo term for the trispectrum also
grows very fast --- typically, like $(P^{1h})^3$ --- and should,
therefore, contribute an important source of noise 
for the power spectrum covariance in the limit of high bias.

When employing a particular cosmological model below, 
we used a standard flat $\Lambda$CDM scenario, with 
$\Omega_m = 0.26$, $n_s = 0.96$ and $\sigma_8 = 0.78$.

\section{Formalism and analytical approximations}

\subsection{The Halo Model}

Over time, gravity enhances the density contrast field by attracting matter towards 
the density peaks, and by creating voids where the density was initially 
below-average. The Halo Model \citep{CooraySheth} describes how this 
process depends on the mass of the collapsed structures by determining, e.g., 
the abundances of the DM halos --- i.e., the mass function, ${d \, \bar{n}_h}/{d \, \log M}$.

Halos are ultimately determined by the peaks of the initial density field, and
according to the theory of peak statistics \citep{BBKS,MoWhite1996}, the main driver of the
abundance of peaks as a function of mass is the mass variance within a certain
comoving radius $R$. The variance of the linear density field inside a spherical top-hat region of 
radius $R$ is determined by:
\bea
\label{Def:sigma}
\sigma^2(R) &=& \frac{1}{2\pi^2}\int dk \, k^2 P_m(k) W(kR) 
\\ \nonumber
&=&\int d\ln k \, 
\Delta_m^2(k)  W(kR) \; ,
\eea
where $P_m(k)$ is the linear matter power spectrum and $W(x) = [3 j_1(x)/x]^2$ is the 
window function for a spherical top-hat region.
The mass contained within radius $R$ at the mean 
background matter density today, $\bar{\rho}_m$, is $M(R) = 4\pi R^3 \bar{\rho}_m/3$.
The peak height is defined as $\nu(M)=\delta_c/\sigma(M)$, where $\delta_c$ is the
linearly extrapolated critical density contrast for spherical collapse.

Due to the statistics of density peaks \citep{BBKS}, all mass functions exhibit an
exponential dependence on the peak height $\nu(M)$. In fact, we have:
\be
\frac{d\bar{n}_h}{d\ln M} = \frac{\bar{\rho}_m}{M} 
\frac{d \ln \sigma^{-1}}{d \ln M} \, f(\nu) \; ,
\ee
where, up to some model-dependent factors and coefficients, $f(\nu) \sim e^{-\nu^2/2}$.

Another crucial ingredient of the Halo Model is the halo bias. Since we would like to
substitute the true matter density contrast $\delta_m $ by the contrast of halo 
counts, $\delta_h = n_h/\bar{n}_h -1$, a relationship between the two 
must be established. We can write this in terms of a local ansatz such as
\citep{FryGaz}:
\be
\delta_h = b_h \delta_m + b_h^{(2)} (\delta_m^2 - \sigma_m^2) + \cdots \, .
\ee
In this paper we will only consider the first term in this relation 
--- the higher-order terms can also become important precisely in the 
limit that we are investigating (tracers with high bias), but we leave 
this key issue for future investigations. 
Typically, the halo bias is a 
smooth power-law function of peak height, such that more massive halos
correspond to higher (and rarer) peaks and have higher values of the 
halo bias.

Our analytical calculations were performed using both the Press-Schechter 
\citep{PS} (PS) 
as well as the 
Sheth-Tormen \citep{ST,SMT} (ST) prescriptions for the mass function 
and halo bias. The semi-analytical calculations of Sec. \ref{Sec:semi-analytic} 
were performed using the Tinker mass function \citep{Tinker} and halo bias 
\citep{Tinker10}. The main results are very similar on all cases.

\subsection{Halo Profile}

Related to the Halo Model, but still a slightly orthogonal result which depends 
more strongly on the non-linear regime of structure formation, is the 
shape of the halo density profile. 
Although our results are completely insensitive to the fine details of
these profiles, in our semi-analytical calculations we employ the standard 
results of \citet{NFW}. In that case, the density profile for a halo of mass $M$ is given by:
\be
\label{Eq:NFW}
\rho(r|M) \sim u(r|M) \sim \frac{1}{\frac{r}{r_s} 
\left( 1 + \frac{r}{r_s} \right)^2} \; ,
\ee
where $r_s=r_s(M)$ is the characteristic scale (``knee'') for a halo of 
mass $M$.
The mass-averaged halo profile in Fourier space is computed as:
\be
u(k|M) = \frac{4\pi}{M} \int dr \, r^2 \, \frac{\sin{kr}}{kr} \, \rho(r|M) \; .
\ee
The important feature for our purpose is that  
$u(k|M) \to 1$ when $k \lesssim 1 \, h$ Mpc$^{-1}$ 
for the range of masses that we are interested in.
This property follows simply from the fact that 
$u(k\rightarrow 0) = (1/M) \int d^3 x \, \rho =1$, and in the range 
$k \lesssim 1 \, h$ Mpc$^{-1}$ this approximation holds for the 
mass scales we are interested in.
Hence our calculations are insensitive to the precise adopted profile, and importantly to whether or not the galaxy count profile is identical to the DM density profile. Note indeed that a discrepancy between these two profiles may have been observed in the innermost regions of the DM halos, where the galaxy profiles may be significantly steeper than the NFW profile \citep{Watson2010,KO2012,Piscionere2014}.

\subsection{Halo Occupation Distribution}

It has been known for a long time that there must be a direct relation between the 
distribution of galaxies and that of the underlying dark matter halos
\citep{WF1991,Kauffmann1993,Navarro1995,MoWhite1996,Kauffmann1999,Springel2005}.
Halo Occupation Distribution (HOD) models provide the probability distribution 
function $P(N|M)$ for a certain number ($N$) of 
galaxies to occupy halos of a given mass ($M$)
\citep{MaFry2000,SeljakHOD,CooraySheth,MartinezBook,Berlind2002,Zheng2005}. 
Although halo mass is not the only factor which determines the number 
of galaxies \citep{Zentner2014},
HODs have been extremely useful to interpret measurements of the 
clustering of different types of
tracers, from galaxies \citep{Zheng2007,Zheng2009} 
to quasars \citep{Porciani,Shen2007,Wake2008,Shen2010,KO2012,Richardson2012}
--- however, in the latter case the simplest HODs may be inadequate to capture the complex
interactions between quasars and their environments 
\citep{Shen2013,Chatterjee2013,CS2015}, and additional parameters such as 
assembly bias should be included.
Often, instead of $P(N|M)$, what is provided are the momenta of the HOD, 
such as $\bar{N}(M) = \langle N \rangle_M $, $ \langle N(N-1) \rangle_M$, etc. 
The brackets define averages over 
halos of the same mass, and the HOD can be defined in terms of these 
momenta. For brevity, we will drop the subscript $M$ from now on.

It is clear that, for very massive halos, the number of galaxies should scale proportionally to the halo mass, but as we approach the low-mass end the situation can be more nuanced.
According to the hierarchical scenario of structure formation, a galaxy can either form inside
its original halo, or join after formation an already existing and populated halo. Such a dichotomy is also seen in numerical simulations \citep{Kravstov2004} and leads to the distinction between ``central'' galaxies, of which there
can be only one per halo, and possibly numerous ``satellite'' galaxies.
A popular functional form for the number of central and satellite galaxies is 
(see, e.g., \citet{Zheng2005}): 
\bea
\label{HOD1}
\langle N_c \rangle &=& \bar{N}_c = \frac12 \, {\rm Erfc} \, \left( \frac{M_c - M}{\sqrt{2} \, \sigma_g} \right) 
\\
\label{HOD2}
\langle N_s \rangle &=& \bar{N}_s = \bar{N}_c \times \tilde{N}_s \; ,
\eea
where:
\be
\label{HOD3}
\tilde{N}_s  = \theta(M-\kappa_g M_c) \, \left( \frac{M - \kappa_g M_c}{M_1} \right)^\alpha \; .
\ee
As denoted by Eq. (\ref{HOD2}), the existence of satellites is  
conditional on the existence of a central galaxy.
Typical values for the HOD parameters are  
$M_c \simeq 10^{13.5} \, h^{-1} \, M_\odot$, 
$M_1 \simeq 10^{14} \, h^{-1} \, M_\odot$, $\alpha \simeq 0.9-1.0$, 
$\kappa_g \simeq 1.1$, $\sigma_g \simeq 1$ \citep{Zheng2005} ---
although, in the case of quasars, especially at high redhifts, some 
parameters can deviate significantly from these values \citep{Chatterjee2013}.

Besides the mean numbers (or richness), we must also specify the higher-order momenta of $P(N,M)$.
If we are only interested in the 2-halo and in the 1-halo terms, then all we need are the
expectation values $\langle N_c^2 \rangle$, $\langle  N_c N_s \rangle$ and $\langle N_s^2 \rangle$. 
The model separating central and satellite galaxies naturally provides these momenta.
By definition, the central galaxy either exists ($N_c=1$) or does not exist ($N_c=0$) inside a halo  
so $\langle N_c(N_c-1)\rangle = 0$ or equivalently $\langle N_c^2 \rangle = \bar{N}_c$. 
Regarding the cross-correlation between central
satellite galaxies, notice that satellites can only exist if there is already at least one central galaxy, 
so $\langle  N_c N_s \rangle = \bar{N}_s$.
As for the satellites, we can assume a simple Poisson distribution, which means, in particular, 
that $\langle N_s^2 \rangle =  \bar{N}_s ( \tilde{N}_s + 1)$.

\subsection{Combining the Halo Model and the HOD}
\label{Subsec:Comb-HMnHOD}

The Halo Model allows us to compute several quantities of interest 
from these ingredients. The mean galaxy number density is:
\be
\label{Def:barn}
{\bar{n}}_g = \int d\ln M \frac{d\bar{n}_h}{d\ln M} \times \bar{N}(M) \; ,
\ee
where $\bar{N} = \bar{N}_c + \bar{N}_s$. 
The galaxy bias is given by:
\be
\label{Def:bg}
b_g (k) = \frac{1}{\bar{n}_g}\int d\ln M \frac{d\bar{n}_h}{d\ln M}  \times \bar{N}(M) b(M) u(k|M) \; ,
\ee
where recall that $u(k|M) \to 1$ for $k \lesssim 1 \, h$ Mpc$^{-1}$.
In terms of the galaxy bias, the two-halo galaxy power spectrum is given 
by:
\be
P^{2h} (k) = b_g^2(k) P_m (k) \; .
\ee
The 1-halo term, on the other hand, is given by the correlation 
of two {\em different} galaxies in the {\em same} halo:
\bea
\nonumber
P^{1h} (k) 
&=& \frac{1}{\bar{n}_g^2}\int d\ln M \frac{d\bar{n}_h}{d\ln M}  \times \langle N(N-1) \rangle \, \left| u(k|M) \right|^2 
\\ \label{Def:P1h}
&=& \frac{1}{\bar{n}_g^2}\int d\ln M \frac{d\bar{n}_h}{d\ln M}  
\\ \nonumber
& & \times 
\bar{N}_c (M) \, [ 2 \tilde{N}_s(M) + \tilde{N}_s^2(M) ] \, 
\left| u(k|M) \right|^2\; ,
\eea
where the term inside square brackets in the third line
is the intra-halo number variance, given the assumptions 
outlined above. 
As usual, the 1-halo term of the power spectrum does not include
the contribution arising from self-correlations of galaxies with themselves 
(i.e., shot noise): in fact, when there is a single galaxy in a halo, 
that is, by definition, the central galaxy, so the number of 
satellites is zero\footnote{Notice that the one-halo terms vanish in the absence of satellite galaxies.
However, due to the Poissonian nature of the HOD, it is possible for a halo to have
satellites even if $\langle N \rangle =1$ for the mass of that halo.}
The measured galaxy power spectrum is therefore:
\be
\label{TotP}
P_g = P^{2h} + P^{1h} + P_S \; ,
\ee
where $P_S$ takes into account the shot noise power spectrum --- which, 
under the assumption of Poissonian statistics for the galaxy counts, is 
given by $P_S = 1/\bar{n}_g$.

Similar arguments can also be applied to higher-order
correlations. The trispectrum $T(\vec{k}_1,\vec{k}_2,\vec{k}_3,\vec{k}_4)$, i.e. the 4-point function in Fourier space, is of particular interest as it determines the covariance of the power spectrum, in its limit $T(\vec{k},-\vec{k},\vec{k}',-\vec{k}')$.
The 1-halo term of this part of the trispectrum is given by:
\bea
\label{Def:T1h}
T^{1h}_g (\vec{k},-\vec{k},\vec{k}',-\vec{k}') 
&=& \frac{1}{\bar{n}_g^4}\int d\ln M \frac{d\bar{n}_h}{d\ln M}  
\\ \nonumber
& & \times \langle N(N-1)(N-2)(N-3) \rangle 
\\ \nonumber
& & \times \left| u(k|M) \right|^2 
\left| u(k'|M) \right|^2 
\\ \nonumber
&=& \frac{1}{\bar{n}_g^4}\int d\ln M \frac{d\bar{n}_h}{d\ln M}  
\\ \nonumber
& & \times 
\bar{N}_c [4 \tilde{N}_s^3 +  \tilde{N}_s^4] 
\\ \nonumber
& & \times  
\left| u(k|M) \right|^2 \left| u(k'|M) \right|^2
\; ,
\eea
where on the second line of the equation above we used the 
same assumptions about the statistics of central and satellite 
galaxies that were used to obtain the expression in
the second line of Eq. (\ref{Def:P1h}).

For our purposes, we will consider scales larger than the size of
the largest halos, so we take $u(k|M) \to 1$ in all our expressions from now on.
We have checked that this is an excellent approximation for 
$k \lesssim 1 \, h$ Mpc$^{-1}$.


\subsection{Analytical model for the peak height}
\label{Subsec:PH}

In some of the following subsections we will perform analytical computations of 
several quantities of interest in the PS \citep{PS} and ST \citep{ST,SMT} formalisms. 
In order to carry out those calculations we need an analytical approximation for the 
peak height in terms of the halo mass. 

The variance of the linear density field inside a tophat spherical region of radius 
$R$ was given in Eq. (\ref{Def:sigma}).
Since the spherical tophat window function has the features that $W(0)=1$ and 
$W \rightarrow 0$ for $x\gg 1$, with a full width at half maximum 
of approximately $x_{FWHM} \approx 1$, it is fair to approximate the variance as:
\bea
\sigma^2(R) \approx \Delta_m^2(k=k_R) \propto k_R^3 \, P_m(k_R) \; ,
\eea
where $k_R = 1/R$.
For self-similar models in which $P_m(k)\propto k^n$, we have $\sigma^2(R)\propto k_R^{n+3} \propto R^{-(n+3)}$. In terms of 
the mass contained inside radius $R$ at the mean background density, $M \propto R^3$, we have:
\bea
\sigma^2(M) \propto M^{-(n+3)/3} \; .
\label{Eq:sigma_M}
\eea
Even though LSS in a standard $\Lambda$CDM Universe is not described by a 
self-similar model, it will be useful to consider this case as it will allow us to obtain interesting analytical expressions for fixed $n$.

\begin{figure}
	\begin{center}
    \includegraphics[width=8cm]{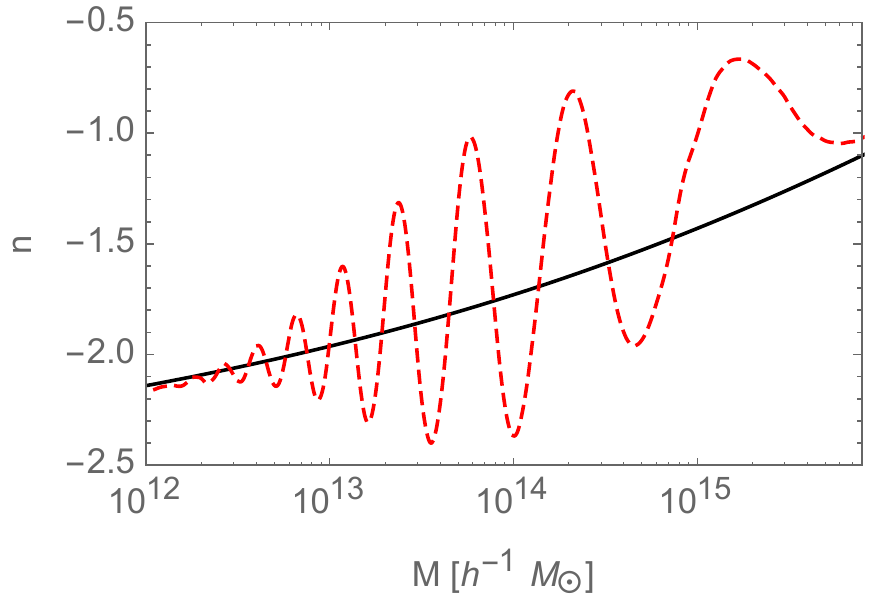}
    \caption{Power-law index as a function of $M$. 
    Solid (black) line: $n = -3(1+d \ln \sigma^2 /d \ln M)$. 
    Dashed (red) line: $n_P = d \ln P(k)/d \ln k$,
    evaluated at $k=k_R=1/R(M)$.
    The wiggles seen in $n_P$ are caused by the BAOs.}
    \label{Fig:sigma_M_relation}
    \end{center}
\end{figure}

The peak height, normalized to 1 at $M=M^*$, is given by:
\bea
\label{Eq:nuap}
\nu=\left(M/M_*\right)^{(n+3)/6} \; ,
\eea
where $M_*$ is typically $\sim 2. \times 10^{13} \, h^{-1} \, M_{\odot}$ in the 
$\Lambda$CDM models.
In particular, within this approximation we have 
$d \ln \sigma^{-1} / d \ln M = (n+3)/6$.

We check the approximation of Eq. (\ref{Eq:sigma_M}) 
in Fig.~\ref{Fig:sigma_M_relation}, where we plot the power-law index 
$n = -3(1+d \ln \sigma^2/d \ln M)$ (solid, black line) together with
$n_P = d \ln P(k)/d \ln k$, evaluated at $k = k_R= 1/R(M)$ (dashed, red line).
The slope of the power spectrum, $n_P$, shows the wiggles from the BAOs. 
The power index of the peak height, on the other hand, is an average over 
several different scales, hence it is 
only sensitive to the mean slope of the power spectrum.
It is clear that the two are closely related, and that the approximation of
Eq. (\ref{Eq:nuap}) holds quite well for $n$ between $-2$ and $-1$ in 
the mass range considered. 
In Sections~\ref{Sec:semi-analytic} and~\ref{Sec:simu} we do not use this 
approximation anymore, and instead compute $\sigma(M)$ from the power 
spectrum in a $\Lambda$CDM model.

\subsection{Mass functions and halo bias}
\label{Subsec:MFs}

The simplest case is that of the Press-Schechter (PS) formalism \citep{PS}.
It provides closed-form expressions for the mass function and halo bias:
\bea
\label{Eq:fps}
f_{PS} (\nu) &=& \sqrt{\frac{2}{\pi}}\nu \exp{[-\nu^2/2]} \; ,
\\ \label{Eq:bps}
b_{PS} (\nu) &=& 1+\frac{\nu^2-1}{\delta_c} \; .
\eea
These formulas are in poor agreement with
the data and N-body simulations, however, when used in
conjunction with an extremely simple HOD, they yield 
simple, straightforward analytical calculations whose 
results convey the basic message of this paper --- 
see Section \ref{Subsec:PS}.

A better fit to simulations and observations is given by the ST mass function and halo bias \citep{ST,SMT}:
\bea
f_{ST} &=& A \sqrt{\frac{2a}{\pi}} \left[ 1 + (a\nu^2)^{-p} \right] \, \nu \, e^{-a\nu^2/2} \; ,
\\
b_{ST} & = & 1 + \frac{a\nu^2-1}{\delta_c} + \frac{2p}{\delta_c [ 1 + (a\nu^2)^p ] } \; ,
\eea
where $A \simeq 0.322$, $a\simeq 0.71$, and $p\simeq 0.3$.
The ST framework gives a more accurate description compared
to PS, while still allowing for fully analytical calculations.
In Section \ref{Subsec:ST} we use the ST formulas and 
a slightly more realistic HOD compared with the calculation
in the PS case --- yet the main results of that Section 
are basically unchanged.

Finally, we also consider the expressions found by \citet{Tinker, Tinker10}, 
which were calibrated from numerical simulations.
The mass function \citep{Tinker} and bias \citep{Tinker10} are given in this case by:
\bea
\label{Tinker}
f(\sigma) &=& 0.186 \times \left[ 
\left( \frac{\sigma}{2.57} \right)^{-1.47} + e^{-1.2/\sigma^2} 
\right] \; , \\
\label{tb}
b_h(\nu) &=& 1 - A_T \frac{\nu^{a_T}}{\nu^{a_T}
+\delta_c^{a_T}} + B_T \nu^{b_T} + C_T \nu^{c_T} \; ,
\eea
where $A_T = 1+0.24 \, y \, \exp [-(4/y)^4]$ 
(with $y=\log_{10} \Delta$, where we choose
$\Delta = 200$), $a_T = 0.44 \, y - 0.88$, $B_T = 0.183$, $b_T = 1.5$, 
$C_T = 0.019 + 0.107 \, y + 0.19 \, \exp [-(4/y)^4]$,
and $c_T = 2.4$.
We will employ this mass function in our semi-analytical calculations,
assuming now a realistic HOD --- see Section \ref{Sec:semi-analytic}. 
As we shall see shortly,  the results are qualitatively identical to 
those of Sections \ref{Subsec:PS} and \ref{Subsec:ST}, which were
found by means of analytical calculations.

\section{Applications}
\label{Sec:Apps}

\subsection{Press-Schechter mass function and a simple HOD}
\label{Subsec:PS}

We begin assuming an extremely simplified HOD, which should hold in an approximate sense 
for sufficiently high halo masses (see, e.g., \citet{Porciani}):
\bea
\label{HOD_simple}
\bar{N}(M)=  \left( \frac{M}{M_1} \right)^\alpha \, \theta(M-M_1) \; ,
\eea
where $\theta(x)$ is the Heaviside step-function.
In this simple HOD we take the cut-off mass to be equal to the mass scale $M_1$.
This HOD also assumes that all galaxies are satellites.
In the final subsections we recover the full description in terms of $\bar{N}_c$ and $\bar{N}_s$, and show
that the central galaxies are unimportant in the limit we are interested in ($b_g \gtrsim 3$).
Except for the low-mass limit, the halo richness should scale roughly proportional to its 
mass, so $\alpha \approx 1$. 
We will assume for the moment that $M_1$ also defines the threshold for finding galaxies in halos -- i.e., 
$\bar{N} = 0$ for $M<M_1$. This approximation 
will be improved in the next subsection, where we carry out the same calculations as here, but using the 
Sheth-Tormen formalism. As we will see, this does not change significantly our main results.

We start by computing the number density of halos which host at least one 
galaxy in our simple HOD, Eq. (\ref{HOD_simple}). Since the number of galaxies 
in each halo follows a Poisson distribution, this is given by:
\be
\label{nhg}
\bar{n}_{h/g} = \int d \ln M \frac{d \bar{n}_h}{d \ln M} \left[ 1-\exp(-\bar{N}) \right]
\ee
Another definition, which will become more useful later on, is the number of halos 
that \emph{could} contain galaxies:
\bea
\label{Eq:nhgps}
\bar{n}_{h,g} &=& \int_{M_1}^\infty d \ln M \frac{d \bar{n}_h}{d \ln M} 
\\ \nonumber
& = & \int_{M_1}^\infty d \ln M \, \frac{\rho_m}{M} \frac{d \ln \sigma^{-1}}{d \ln M} \sqrt{\frac{2}{\pi}} \nu \exp{[-\nu^2/2]}
\\ \nonumber
& = & \sqrt{\frac{2}{\pi}}\frac{\rho_m}{M_*} \int_{\nu_1}^\infty d \nu \, \nu^{-6/(n+3)} e^{-\nu^2/2} \; , 
\eea
where we have used the PS mass function and the approximations 
$M=M_* \nu^{6/(n+3)}$, as well as the definition $\nu_1 = (M_1/M_*)^{(n+3)/6}$.\\
For high $M_1$ the integral in Eq. (\ref{nhg}) is dominated by the exponential behavior of 
the mass function, and we can replace the $\bar{N}$ in the exponent by the mean 
number of galaxies in the halos just above the cut-off mass scale,
$\bar{N}(M) \to \bar{N} (M_1) = \bar{N}_{\rm min}$. 
Hence, the actual number of halos containing galaxies 
can be approximated by $\bar{n}_{h/g} \approx (1-e^{-\bar{N}_{\rm min}})\times \bar{n}_{h,g}$. 
For the HOD of Eq. (\ref{HOD_simple}) this minimum mean number of galaxies is $\bar{N}_{\rm min} =1$, so $\bar{n}_{h/g} \approx 0.63 \; \bar{n}_{h,g}$.

With the variable change $x\equiv \nu^2/2$ we obtain:
\be
\bar{n}_{h,g} = \sqrt{\frac{2}{\pi}}\frac{\rho_m}{M_*} 2^{\lambda_0} \int_{x_1}^\infty dx \, x^{\lambda_0} e^{-x} 
= \sqrt{\frac{2}{\pi}}\frac{\rho_m}{M_*} 2^{\lambda_0} \Gamma(1+\lambda_0,x_1)
\; ,
\ee
where $\lambda_0 = -1/2 - 3/(n+3)$, $x_1 = \nu_1^2/2$, and $\Gamma (\kappa,x)$ 
is the (upper) incomplete Gamma function of order $\kappa$. 
Typically, $-2 \lesssim n \lesssim -1$ for halos at the scales of interest,
which means that $-7/2 \lesssim \lambda_0 \lesssim -2$.

The incomplete Gamma function is related to the simple Gamma function by 
$\Gamma(\kappa) = \Gamma(\kappa,x=0)$, and has asymptotic limits given by:
\be
\lim_{x\to0} \Gamma (\kappa,x) \; \rightarrow  \; \Gamma(\kappa) - x^{\kappa} 
\left[ \frac{1}{\kappa} - \frac{x}{1+\kappa} + {\cal{O}}(x^2) \right]
\; ,
\ee
and
\be
\label{ExpGamma}
\lim_{x\to\infty} \Gamma (\kappa,x) \; \rightarrow  \; e^{-x} \, x^{\kappa-1} \left[ 1
+ \frac{\kappa-1}{x} + {\cal{O}}(x^{-2}) \right]
\; .
\ee
Notice, in particular, that $\lim_{x\to\infty} \Gamma (1+\kappa,x)/\Gamma (\kappa,x)  \rightarrow 1
+ x + {\cal{O}}(x^{-1})$. It is also interesting to note that,
for $0 \lesssim \kappa \lesssim 1 $, the asymptotic expression of Eq. (\ref{ExpGamma}) 
is remarkably accurate down to $x \simeq 1$.

Hence, for the ranges of interest for $n$, in which $\lambda_0$ is negative,  
the number density of the halos that host galaxies 
should diverge in the limit  $x_1 \rightarrow 0$ --- i.e., when $M_1 \ll M_*$.
Indeed, the number of halos of arbitrarily small masses is 
arbitrarily large, unless we specify a 
smoothing scale $R_f$, in which case it asymptotes 
to $\bar{n}_{h,g} \propto R_f^{-3}$ \citep{BBKS}.

Similarly as was done above, we can compute analytically the quantities defined in Section \ref{Subsec:Comb-HMnHOD}. 
For the mean number density of galaxies we obtain:
\bea
\label{Eq:ngps}
\bar{n}_g &=& \int_{M_1}^\infty d\ln M \, \frac{d\bar{n}_h}{d\ln M} \times \bar{N}(M)  
\\ \nonumber
&=& \sqrt{\frac{2}{\pi}} \frac{\rho_m}{M_*} \left( \frac{M_*}{M_1} \right)^\alpha 
2^{\lambda_1} \Gamma(1+\lambda_1, x_1) \; ,
\eea
where we have used the same definitions as above, 
with the difference that now the index is
$\lambda_1 = \lambda_0 + 3\alpha/(n+3) = -1/2 + 3(\alpha-1)/(n+3)$. 
Since $\alpha \simeq 1$ and $-2 \lesssim n \lesssim -1$,
we have $ \lambda_1 \simeq -1/2 $.
Notice that for the case $\alpha=1$, $\lambda_1=-1/2$ and 
in the limit $x_1\rightarrow 0$, we have
$\bar{n}_g = \rho_m/M_1 = \rho_m\langle N \rangle /M$

Interestingly, using Eq. (\ref{ExpGamma}) we find that in the high mass 
limit ($x_1\rightarrow \infty$), $\bar{n}_g = \bar{n}_{h,g}$ 
-- i.e., in that case the number of halos that could host a galaxy 
is equal to the mean number of galaxies.
This is a consequence of the simple HOD, Eq. (\ref{HOD_simple}), which takes
the cut-off mass to be identical to the mass scale $M_1$.
Since the galaxy bias increases with $M_1$, for high values of this mass 
the number of halos above the cut-off is exponentially suppressed, and only
the least massive halos are populated with galaxies. 
In this case each halo ends up hosting only one galaxy --- or,
more accurately, because of Poisson statistics, about 63\% of halos contain 
only one galaxy, and the rest contain two or more galaxies.

We can calculate the galaxy bias in the same fashion, using the 
PS halo bias of Eq. (\ref{Eq:bps}):
\be
b_g = 1-\delta_c^{-1} + 2 \delta_c^{-1} 
\frac{\Gamma(2+\lambda_1,x_1)}{\Gamma(1+\lambda_1,x_1)} \; .
\ee
In the $x_1 \rightarrow 0$ limit we can use the property of the Gamma function
$\Gamma(2 + \lambda_1)=(1 + \lambda_1) \Gamma(1 + \lambda_1)$,
which leads to $b_g \simeq 1-\delta_c^{-1} + 2 \delta_c^{-1} (1+\lambda_1)$. 
We also note that taking $\alpha=1$ leads to $\lambda_1=-1/2$ and $b_g=1$, 
simply reflecting the halo bias consistency relation.
On the other hand, in the limit of very large threshold masses ($x_1 \gg 1$) we obtain
$b_g \simeq 1+ \delta_c^{-1} + 2 \delta_c^{-1} \, x_1$.
Hence, in order to increase bias, it is sufficient that 
$x_1 \gg 1$ --- however, this is not a necessary condition: one could also fix 
the cut-off mass scale and decrease $n$, or increase $\alpha$.

A similar calculation as the one performed in Eq. (\ref{Eq:ngps})
leads to an expression for the 1-halo term of the 
galaxy power spectrum:
\bea
\nonumber
P^{1h} &=& \frac{1}{\bar{n}_g^2}
\int d\ln M \, \frac{d\bar{n}_h} {d\ln M} \bar{N}^2
\\ \label{P1h0}
&=& 
\left[ \sqrt{\frac{2}{\pi}} \frac{\rho_m}{M_*} 2^{2\lambda_1-\lambda_2}\right]^{-1}  
\times \frac{ \Gamma(1+\lambda_2,x_1) }{[\Gamma(1+\lambda_1,x_1)]^2} \; ,
\eea
where on the first line the term $\langle N(N-1) \rangle$ reduces to $\bar{N}^2$ since we only have satellite galaxies with a Poisson distribution, and on the second line
we have substituted the expression for $\bar{n}_g$ and $\lambda_2 = \lambda_0 + 6 \alpha/(n+3)$.
Using the expression for the number density of halos hosting these galaxies, Eq. (\ref{Eq:nhgps}),
we obtain:
\be
\label{P1h}
P^{1h} = 
\frac{\Gamma(1+\lambda_0,x_1) \, \Gamma(1+\lambda_2,x_1) }{[\Gamma(1+\lambda_1,x_1)]^2}
\times \frac{1}{\bar{n}_{h,g}} \; .
\ee

Notice that the three indices, $\lambda_0$ (that appeared in $\bar{n}_{h,g}$), $\lambda_1$
(which appeared in $\bar{n}_g$ and $b_g$), and $\lambda_2$, are related by $\lambda_0 = 2 \lambda_1 - \lambda_2$. By substituting the asymptotic expression for $\Gamma(\kappa,z)$ in the limit 
$z \rightarrow \infty$ one can verify that the prefactor appearing Eq. (\ref{P1h}) is
$ \Gamma(1+\lambda_0,x_1)  \Gamma(1+\lambda_1,x_1)/ [\Gamma(1+\lambda_1,x_1)]^2 \simeq 1$. 
Hence, in the limit of high threshold mass, $P^{1h} \approx 1/\bar{n}_{h,g}$ .

In the opposite limit, of low threshold masses, the prefactor can become quite large --- but then so does
$\bar{n}_{h,g}$ become large. In general, this case corresponds to tracers with low biases.
In that limit, it is convenient to revert to the original expression, Eq. (\ref{P1h0}), and write instead:
\bea
\label{P1h2}
P^{1h} &= &\frac{2^{\lambda_2-\lambda_1}}{\bar{n}_g}
\left(\frac{M_*}{M_1} \right)^\alpha
\frac{\Gamma(1+\lambda_2,x_1)}{\Gamma(1+\lambda_1,x_1)} 
\\ \nonumber
&\simeq& 
\frac{2^{\lambda_2-\lambda_1}}{\bar{n}_g} 
\left(\frac{M_*}{M_1} \right)^\alpha
\frac{\Gamma(1+\lambda_2)}{\Gamma(1+\lambda_1)}
\; .
\eea
Let us suppose that we can change the HOD parameters while maintaining the mean number of galaxies, 
$\bar{n}_g$, fixed. In the limit of low-mass threshold the bias is 
$b_g \approx 1 - \delta_c^{-1} + \delta_c^{-1} [1 + 6(\alpha-1)/(n+3)]$ 
--- i.e., higher values of $\alpha$ 
and/or lower values of $n$ correspond to higher biases. Since $\lambda_2 = \lambda_1 + 3\alpha/(n+3)$,
the ratio of Gamma functions in Eq. (\ref{P1h2}) can be regarded as a steep function of bias.

From Eq. (\ref{P1h}) we see that $P^{1h}$, which is proportional to $1/\bar{n}_{h,g}$, 
plays the role of  a {\it halo} shot noise term.
For highly biased tracers (whose mass thresholds are relatively high), the galaxies end up
in just a few halos, so that the halo shot noise term becomes an important part of the power
spectrum and its covariance. This means, in particular, that the 
barrier for measuring the (2-halo) power spectrum with highly biased tracers may not 
be just the shot noise of that tracer, but also an additional {\em halo shot noise} 
coming from the 1-halo term.

\begin{figure}
		\begin{center}
		\resizebox{8.cm}{!}{\includegraphics{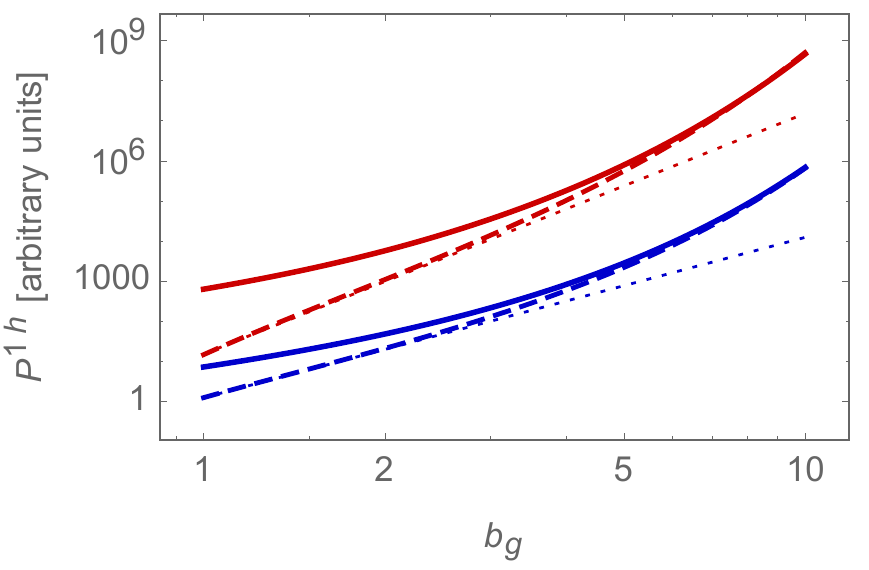}}
		\caption{Scaling of the 1-halo term in the PS model. The
		exact formula, Eq. (\ref{P1h0}), is denoted by the thick solid lines 
		(upper line: $n=-2$; lower line: $n=-1$), while the dashed lines
		correspond to the approximation of Eq. (\ref{Eq:appP1h0}). The dotted
		lines are the power laws $b_g^4$ (for the case $n=-1$) and 
		$b_g^6$ (for the case $n=-2$).}
		\label{Fig:simple_scaling}
		\end{center}
\end{figure}

In the limit of high bias (and high mass thresholds) we can 
neglect the subdominant dependence in the prefactor of $1/\bar{n}_{h,g}$ 
in Eq. (\ref{P1h}), and express the contribution of the halo shot noise term as a 
function of bias. Using 
$b_g \simeq 1 + \delta_c^{-1} + 2 \delta_c^{-1} x_1 \rightarrow 2 \delta_c^{-1} x_1 $, the
mean number density of halos can be expressed as:
\be
\bar{n}_{h,g} 
\sim x_1^{-\frac{n+9}{2(n+3)}} e^{-x_1} 
\sim b_g^{-\frac{n+9}{2(n+3)}} e^{-b_g \, \delta_c/2} \; ,
\ee
hence:
\be
\label{Eq:appP1h0}
P^{1h} \sim x_1^{\frac{n+9}{2(n+3)}} e^{x_1} 
\sim b_g^{\frac{n+9}{2(n+3)}} e^{b_g \, \delta_c /2} \; .
\ee
For intermediate values of the galaxy bias the 1-halo
term is well approximated by a power-law.
In Fig. \ref{Fig:simple_scaling} we show the exact formula, Eq. (\ref{P1h0}), for the cases 
$n=-1$ (lower solid line) and $n=-2$ (upper solid line), in arbitrary units. 
The first approximation in the middle of Eq. (\ref{Eq:appP1h0}) is plotted as the dashed lines for the two cases,
where we used $x_1 \approx (1-\delta_c+\delta_c \, b_g)/2$ .
One can see that the approximation becomes better for higher values of the bias.
Also plotted  (dotted lines) are the power laws $b_g^4$ (for the case $n=-1$) and 
$b_g^6$ (for the case $n=-2$). The simple power laws shown in Fig. \ref{Fig:simple_scaling} 
are a good approximation in the interval $1.5 \lesssim b_g \lesssim 3$.

Since the 2-halo term scales as $P^{2h} = b_g^2 \, P_m$, but 
the 1-halo term grows much faster with bias, the latter component
should become increasingly important for highly-biased tracers. In fact, this 
already happens at small scales ($k \gtrsim \; 1 \; h$ Mpc$^{-1}$) 
even for galaxies with relatively low biases. If we select tracers 
with increasing values of the bias (e.g. quasars), the 1-halo term will become more important, 
even in the large-scale limit, acting effectively as a type of ``halo shot 
noise''. However, in contrast to the usual situation where shot 
noise can be beaten down by observing a larger number of 
galaxies, when the tracers are very highly biased this 
halo shot noise cannot be lowered, and a limiting factor 
for measuring the power spectrum is the finite number of 
halos, and not only the number of galaxies in the survey.

It is easy to see that this argument also applies to higher-order correlation functions. 
The same type of integral computed above appears also in the 1-halo term of the trispectrum,
Eq. (\ref{Def:T1h}), and taking $k \to 0$ and $k' \to 0$ leads to:
\bea
\nonumber
T^{1h} &=& \frac{1}{\bar{n}_g^4}
\int d\ln M \, \frac{d\bar{n}_h} {d\ln M} [ 4 \bar{N}^3 + \bar{N}^4 ]
\\ \nonumber
&=&  \frac{1}{\bar{n}_g^4} \times \sqrt{\frac{2}{\pi}}
\frac{\rho_m}{M_*}
\left[
4 \left(\frac{M_*}{M_1} \right)^{3 \alpha} 
2^{\lambda_3} \Gamma(1+\lambda_3,x_1) \right.
\\ \label{T1h0}
& & \left. +
\left(\frac{M_*}{M_1} \right)^{4 \alpha} 
2^{\lambda_4} \Gamma(1+\lambda_4,x_1)
\right]
\; ,
\eea
where $\lambda_i = \lambda_0 + i \times  3 \alpha/(n+3)$.
In the high-mass, high-bias limit we obtain that:
\be
T^{1h}  \sim b_g^{\frac32 \frac{n+9}{n+3} } e^{3  b_g \, \delta_c/2} \sim \left( P^{1h} \right)^3 \; .
\ee
Thus $T^{1h}$ will become a dominant part of the power spectrum covariance matrix in the high bias limit.\\
The same calculation can be employed generally for the 1-halo term of the $N+1$-th order polyspectrum in the high bias limit, showing that it grows with the scaling $b_g^{\frac{N}{2} \frac{n+9}{n+3}} e^{N b_g \, \delta_c/2} \sim \left(P^{1h}\right)^N$. 

Hence, we conclude that for highly biased tracers
not only the power spectrum, but also the higher-order
statistics, are increasingly affected by intra-halo
statistics, and may become effectively limited not only by the 
counts of the tracers, but by the counts of the halos as well.

\subsection{Sheth-Tormen mass function and a simple HOD}
\label{Subsec:ST}

In this Section we still consider, as before, a simplified HOD which does not 
distinguish between central and satellite galaxies. 
However, we now consider a cut-off mass for the halo richness, $M_{c}$, 
which is different from the mass scale $M_1$ (in fact, typically 
$M_c < M_1$  --- see, e.g., \citet{Tinker}). Hence, our HOD is:
\bea
\label{HOD_simple2}
\bar{N}(M)= \left( \frac{M}{M_1} \right)^\alpha \, \theta(M-M_c) \; .
\eea

Defining the variable:
\be
x = \frac{a}{2} \nu^2 \rightarrow \frac{a}{2} \left( \frac{M}{M_*} \right)^{\frac{n+3}{3}} \; ,
\ee
and the cut-off:
\be
\label{xc}
x_c = \frac{a}{2} \left( \frac{M_c}{M_*} \right)^{\frac{n+3}{3}} \; ,
\ee
the calculations of the previous Section can now be performed in basically 
the same fashion. The number density of halos that can host galaxies is:
\begin{align}
\nonumber
\bar{n}_{h,g} &= \frac{\rho_m}{M_*} \, \frac{A}{\sqrt{\pi}} \, 
\left( \frac{2}{a} \right)^{\lambda_0}  
\int_{x_c}^{\infty}  dx \,  x^{\lambda_0} \left[ 1 + (2x)^{-p} \right] e^{-x}
\\
& =  \bar{n}_0 \left( \frac{2}{a} \right)^{\lambda_0}  
\left[ \Gamma(1+\lambda_0, x_c) + 2^{-p} \Gamma(1+\lambda_0 - p, x_c) \right] \; ,
\end{align}
where, as previously, $\lambda_0 = -1/2 - 3/(3+n)$, and we have defined 
$\bar{n}_0 = \rho_m A /M_* \sqrt{\pi}$. 

\begin{figure}
		\begin{center}
		\resizebox{8.cm}{!}{\includegraphics{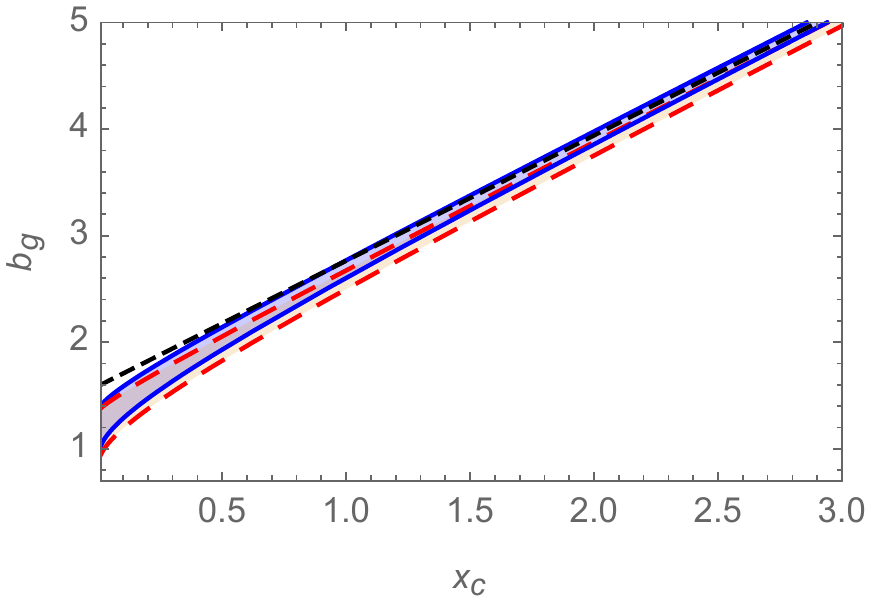}}
		\caption{Galaxy bias obtained using the Sheth-Tormen formalism 
		(blue lines and filled region), and
		in the Press-Schechter formalism (red, long-dashed lines 
        and filled region),
		as a function of the cut-off scale $x_c$ [see Eq. (\ref{xc})].
		The parameters were allowed to range in the intervals 
		$0.9 < \alpha < 1.1$, and $ -2 < n < -1$.
		The black (short-dashed) line shows the approximation of 
        Eq. (\ref{bapp}).}
		\label{Fig:bias}
		\end{center}
\end{figure}

A similar calculation leads to the mean number density of galaxies:
\bea
\bar{n}_g &=& \bar{n}_0 \,  \left( \frac{M_*}{M_1} \right)^\alpha \,
\left( \frac{2}{a} \right)^{\lambda_1}
\\ \nonumber
& & \times 
\left[ \Gamma(1+\lambda_1, x_c) + 2^{-p} \Gamma(1+\lambda_1 - p, x_c) \right] \; ,
\eea
and the galaxy bias becomes:
\bea
b_g &=& 1 - \delta_c^{-1} 
+ \frac{2 \delta_c^{-1}}{\Gamma(1+\lambda_1, x_c) + 2^{-p} \Gamma(1+\lambda_1 - p, x_c)}
\\ \nonumber
& & \times \left[ \Gamma(2+\lambda_1, x_c) + 2^{-p} \Gamma(2+\lambda_1 - p, x_c) \right.
\\ \nonumber
& & \left. + p \, 2^{-p} \Gamma(1+\lambda_1 - p, x_c) \right] \; .
\eea
In the limit of $x_c \gg 1$ the galaxy bias can be considerably simplified, 
in fact:
\be
\label{bapp}
\lim_{x_c =\infty} b_g \rightarrow 1 - \delta_c^{-1} + 2 \delta_c^{-1} x_c \; ,
\ee
which is basically the approximate expression we obtained in the PS case,
where $x_1$ played the role of the cut-off mass scale.
This relationship between the cut-off scale and the bias allows us to write, in
the limit of high biases,
$x_c \simeq \delta_c (b_g -1 + \delta_c^{-1})/2$. This turns out to be a 
fairly good approximation as can be seen from Fig. \ref{Fig:bias}.
Notice that we do not expect the model of Eq. (\ref{HOD_simple2}) to hold for 
low-biased galaxies, when the small-halo mass limit is critical.

After some algebra, the 1-halo term can be expressed in the same way as was done for the PS case:
\be
\label{p1hgst}
P^{1h} = q \, \frac{1}{\bar{n}_{h,g}} \; ,
\ee
where:
\bea
\label{q}
q &=& \frac{ \Gamma(1+\lambda_0, x_c) + 2^{-p} \Gamma(1+\lambda_0 - p, x_c) }
{\left[ \Gamma(1+\lambda_1, x_c) + 2^{-p} \Gamma(1+\lambda_1 - p, x_c) \right]^2} 
\\ \nonumber
& & \times 
\left[ \Gamma(1+\lambda_2, x_c) + 2^{-p} \Gamma(1+\lambda_2 - p, x_c) \right] \; .
\eea
In the limit of $x_c \gg 1$ we can use the series expansion of Eq. (\ref{ExpGamma})
to show that, as in the PS case, the prefactor $q \rightarrow 1$ 
\footnote{In this limit, we see from Eq. (\ref{p1hgst}) that the 1-halo term of 
the power spectrum inherits a dependence on the number density of the
halos that contain at least one galaxy. However, the HOD we used in this Section and in the 
previous one make no distinction between central and satellite galaxies --- in fact,
we have simply used the typical parametrizations used for satellites. Hence, in
this context, $\bar{n}_{h,g}$ should be regarded as the number density of halos
containing {\em more} than one galaxy.} .
In this limit $P^{1h}$ reduces to the following expression:
\be 
\label{Eq:appP1}
P^{1h} \simeq 
\left[\bar{n}_0 \left( \frac{2}{a} \right)^{\lambda_0} x_c^{\lambda_0} 
\left( 1 + 2^{-p} x_c^{-p}\right)
e^{-x_c}  \right]^{-1}
\ee 
with $x_c$ depending on the bias in the following way:
\be 
x_c = \frac{\delta_c}{2}(b_g -1 + \delta_c^{-1}).
\ee
Conversely, in the limit $x_c \ll 1$, we obtain the following 
scaling for the one-halo term:
\be 
P^{1h} \simeq 
\frac{\Gamma(1+\lambda_2)+2^{-p} \Gamma(1+\lambda_2-p)}{\bar{n}_0 \left(\frac{2}{a}\right)^{\lambda_0} \left[\Gamma(1+\lambda_1)+2^{-p}
\Gamma(1+\lambda_1-p)\right]^2} \; .
\ee

Comparing Eq. (\ref{Eq:appP1h0}) and Eq. (\ref{Eq:appP1}), we see that,
for high values of the bias, the 1-halo term in the ST model behaves
in basically the same way as was found for the PS formalism.

\subsection{Semi-analytical model: Tinker mass function and realistic HOD}
\label{Sec:semi-analytic}

The analytical approximations of the previous Sections have 
allowed us to obtain simple expressions for the number 
densities of galaxies and halos, the bias, and the 1-halo terms, 
but we made some strong assumptions --- in particular, about the 
simple scaling of halo richness, about the assumption of self-similarity which fixed the scaling of 
$\nu$ with mass, and about the way in which we cut off the halo richness
below some given mass scale. Although the final results may 
have seemed natural and physically sensible, they could have
been influenced or even driven by these simplifications.

In this Section we argue that these results are robust.
We show this by improving the modeling of the previous Sections in a number of ways: 
first, we calculate the mass variance
$\sigma(M)$ of Eq. (\ref{Def:sigma}) from the power spectrum 
of a vanilla-$\Lambda$CDM model; second, we employ the 
\citet{Tinker} mass function and halo bias of Eqs. (\ref{Tinker})-(\ref{tb}), which are a 
slightly better fit to the N-body simulations compared to the 
PS or ST expressions; and third, we use a class of HODs 
which is inspired and calibrated by observations 
\citep{Zheng2005,Zheng2007,Zheng2009}. 
We also distinguish between central and satellite galaxies 
--- whereas in the preceding Sections we implicitly assumed 
that all galaxies were satellites.

As for the HOD, we have used the formulas of Eqs. 
(\ref{HOD1})-(\ref{HOD2}) for the halo richness of central and 
satellite galaxies.
Typical values for these parameters are 
$M_c \simeq 10^{13.5} \, h^{-1} \, M_\odot$, 
$M_1 \simeq 10^{14.} \, h^{-1} \, M_\odot$, $\alpha \simeq 0.9$, 
$\kappa_g \simeq 1.1$, and $\sigma_g \simeq 1$ \citep{Zheng2009}.

In contrast to the previous Sections, where all calculations could 
be carried out exactly, here we instead compute numerically the 
galaxy number density of Eq. (\ref{Def:barn}), the bias of Eq.
(\ref{Def:bg}), and the 1-halo term of Eq. (\ref{Def:P1h}). In this 
way we can explore basically any point in parameter space, and compute 
the properties of the galaxy models corresponding to those points.

\begin{figure}
		\begin{center}
		\resizebox{\linewidth}{!}{\includegraphics{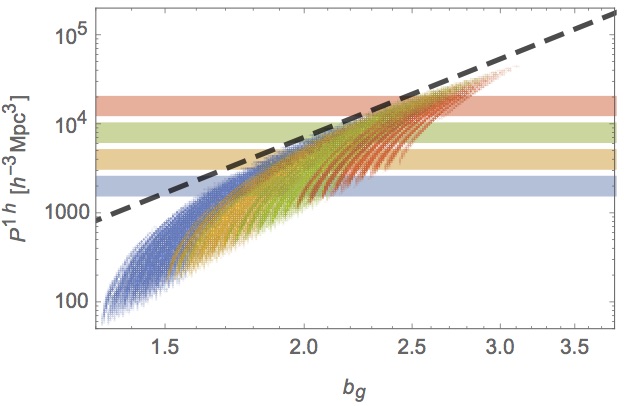}}
		\resizebox{\linewidth}{!}{\includegraphics{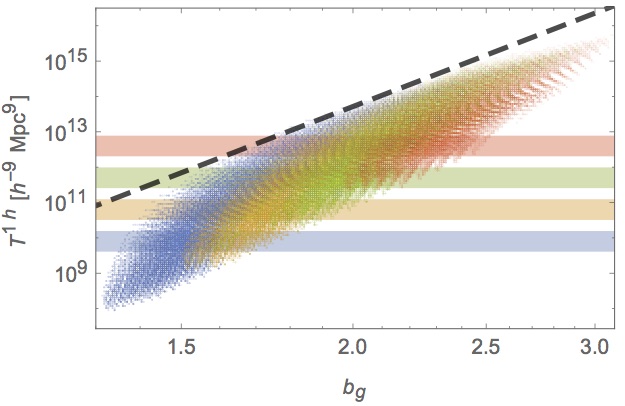}}
		\caption{Top panel: 1-halo term in the HOD of Eqs. (\ref{HOD1})-(\ref{HOD2}). 
		The models were split according to the number density of galaxies.
		From left to right, and from the bottom up, the models shown have 
		$10^{-3.15} \geq \bar{n}_g  \geq 10^{-3.45}$, 
		$10^{-3.45} \geq \bar{n}_g  \geq 10^{-3.75}$, 
		$10^{-3.75} \geq \bar{n}_g  \geq 10^{-4.05}$, 
		and $10^{-4.05} \geq \bar{n}_g  \geq 10^{-4.35}$.
		In each case, the horizontal shaded area marks the 
        corresponding range of $1/\bar{n}_g$.
		The dashed black line is the power-law $b_g^{4.5}$. 
		The stripes seen mostly in the lower right corner 
        are just an artifact of the grid we
		used to explore the HOD parameter space.
		Bottom panel: 1-halo term of the trispectrum for the same class of HODs.
        The dashed black line is the power-law $b_g^{13.5}$. 
        The horizontal shaded areas mark the 
        corresponding ranges of $1/\bar{n}_g^3$.}
		\label{Fig:P1hT1h}
		\end{center}
\end{figure}

We allow the parameters to vary in the following ranges, 
while keeping always $M_c \le M_1$:
\bea
\label{Mcrange}
12.75 & < & \log_{10} M_c \, h/M_\odot  < 14.25 \; ,
\\
\label{M1range}
13.2 & < & \log_{10} M_1 \, h/M_\odot < 15.0 \; ,
\\
0.85 & < & \alpha  < 1.15 \; ,
\\
0.85 & < & \sigma_g  < 1.15 \; ,
\\
1.0 & < & \kappa_g  < 1.3 \; .
\eea
The results for the 1-halo term of the power spectrum are shown in the top panel 
of Fig. \ref{Fig:P1hT1h}.
We have split the different HOD models in four groups, 
according to the number densities of galaxies.
From the bottom, the horizontal shaded areas correspond to
HODs whose ranges of $\bar{n}_g^{-1}$ fall in the intervals
$10^{-L+0.15} \geq \bar{n}_g \, (h^3  \, {\rm Mpc}^{-3}) \geq 10^{-L-0.15}$, 
for $L=3.3$, 3.6, 3.9, and 4.2 .
The values of $P^{1h}$ for models in those four groups are shown as
points of the same colors as the shaded areas, 
from lower left to upper right, respectively.
The dashed black line is the power-law $b_g^{4.5}$.

We saw in Sections \ref{Subsec:PS} and \ref{Subsec:ST} that the one-halo term 
evolves like a power-law in $b_g$ for intermediate values of the galaxy bias. 
This is again what we observe here: the asymptotic behavior of the one-halo follows a steep 
power-law as a function of galaxy bias. 
The growth of the one-halo term is constrained by the requirement that 
$M_c \le M_1$, which limits the number of galaxies in halos, and imposes 
an upper limit on the bias and the one-halo term. 
If we relax this physical requirement, the power-law evolution of the 
one-halo term continues at higher bias. 

The results obtained in Sections \ref{Subsec:PS} and \ref{Subsec:ST} were derived under 
the assumption that all galaxies are of the ``satellite'' type --- an approximation that we 
did not use here.
For highly biased tracers the 1-halo term can be comparable to Poisson
shot noise, which shows that the ``central'' galaxies are less relevant in that limit, 
providing a motivation for that approximation in Sections \ref{Subsec:PS} and \ref{Subsec:ST}.
Conversely, the fact that $P^{1h}$ drops well below the level of galaxy shot noise 
for low values of the bias shows that, for such types of objects, the central 
galaxy plays an important role, as many (or most) halos host a single central
galaxy.

In the bottom panel of Fig. \ref{Fig:P1hT1h} we show the results for the 
1-halo term of the trispectrum of Eq. (\ref{Def:T1h}), in the limit 
$k \to 0$, $k'\to 0$, for the same range of HOD parameters used in the top panel. 
Again, we separate the models in groups, 
according to the number density of galaxies, and the horizontal shaded 
areas denote the different values of $n_g^{-3}$ for each group.
The dashed line is the power-law $b_g^{13.5}$. 

\begin{figure}
		\begin{center}
		\resizebox{\linewidth}{!}{\includegraphics{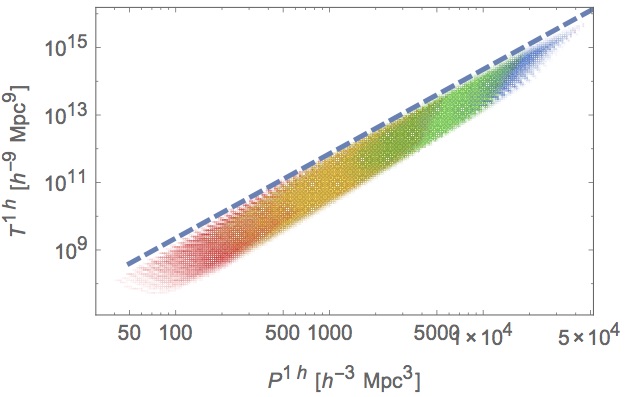}}
		\caption{1-halo term of the trispectrum plotted against the 1-halo term of the
		power spectrum, for the HOD of Eqs. (\ref{HOD1})-(\ref{HOD2}). 
        The dashed black line is the power-law $(P^{1h})^{2.5}$. 
        The models were split according to bias: from lower left to upper right,
        the different colors indicate models with $1.0 \leq b_g \leq 1.5$, 
        $1.5 \leq b_g \leq 2.0$, $2.0 \leq b_g \leq 2.5$,  and $2.5 \leq b_g \leq 3.0$.}
		\label{Fig:T1vP1}
		\end{center}
\end{figure}

As argued above, the 1-halo term of the trispectrum also scales rapidly with bias, 
approximately as $(P^{1h})^3$. In Fig. \ref{Fig:T1vP1} we show the 1-halo term of the trispectrum 
[Eq. (\ref{Def:T1h})] against the 1-halo term of the power spectrum. The dashed line indicates
the scaling $(P^{1h})^{2.5}$.

\subsection{Simulations}
\label{Sec:simu}

We further test the results from the previous sections in a fully
numerical set-up, by using a catalog of halos from the DEUS simulations
\footnote{http://www.deus-consortium.org}
\citep{Alimi2010,Rasera2010,Courtin2011}. We use halos detected with the Friends-of-Friends
algorithm in a box of $648~h^{-1}$~Mpc, containing $1024^3$ dark matter
particles, in a $\Lambda$CDM cosmology in agreement with WMAP5 \citep{WMAP5}.
We should note that the DEUS mass function is very well fit by the Tinker mass function, and
that the DEUS halo bias is also well fit by the Tinker halo bias.


We populate the halos using the HOD formalism presented in Eqs.~(\ref{HOD1}) 
and~(\ref{HOD2}), with fixed parameters $\alpha = 0.9$, $\sigma_g = 1.0$ and 
$\kappa_g = 1.1$. The $M_c$ and $M_1$ parameters are allowed to vary in the 
ranges given in Eqs. (\ref{Mcrange})-(\ref{M1range}).

The number $N_c$ of central galaxies is either 0 or 1, and is drawn from a nearest-integer 
uniform distribution with mean $\bar{N}_c$.
For halos containing a central galaxy, the number of satellite galaxies is then drawn 
from a Poisson distribution with mean $\tilde{N}_s$ (since we are taking the 
$k \to 0$ limit for the 1-halo term, it is irrelevant where in the halo those satellite
galaxies are placed). Obviously, halos that do not contain a central galaxy have no 
satellite galaxies.

With the DEUS halo catalog, and the HOD implementation described above, expressions
such as that for the number density of galaxies become:
\bea
\label{Eq:ngpsSIM}
\bar{n}_g &=& \int_{M_1}^\infty d\ln M \, \frac{d\bar{n}_h}{d\ln M} \times \bar{N}(M)  
\\ \nonumber
& \longrightarrow &
\frac{1}{V} \sum_h \, \left\{ N_c[M(h)] + N_s [M(h)] \right\} 
\\ \nonumber
&=&
\frac{1}{V} \sum_M \, N_h (M) \, [\bar{N}_c(M) + \bar{N}_s(M) ] \; ,
\eea
where $V$ is the volume of the simulation. In order to compute galaxy bias, 
for simplicity we employ the Tinker bias, Eq. (\ref{tb}), in Eq. (\ref{Def:bg}).

\begin{figure}
		\begin{center}
        \includegraphics[width=8.7cm]{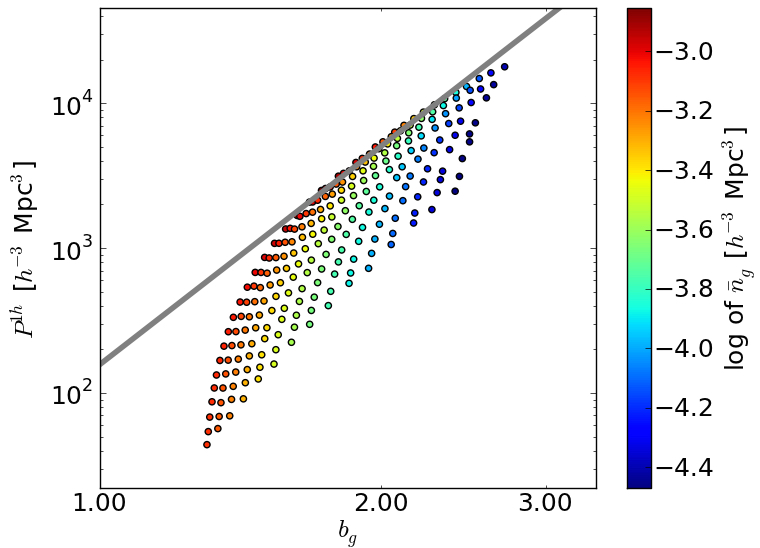}
        \includegraphics[width=8.5cm]{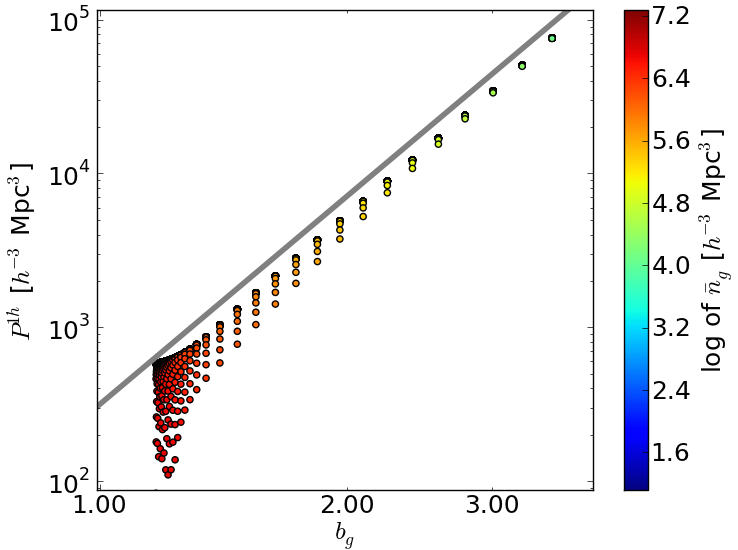}
        \caption{One-halo term and galaxy bias obtained from 
        the DEUS simulated halo catalog populated with the HOD 
        of Eqs. (\ref{HOD1})-(\ref{HOD2}) (top panel), and with 
        the test HOD of Eq. (\ref{Eq:AltHOD_1}) (bottom panel).
        In the top panel, the grey line shows the scaling $b_g^{5}$, 
        while in the bottom panel the scaling is $b_g^{4.5}$.}
        \label{Fig:Bias_HOD_simu}
        \end{center}
\end{figure}

The resulting scaling between the one-halo term and the galaxy bias is shown 
in the top panel of Fig.~\ref{Fig:Bias_HOD_simu}. The same type of scaling found in the 
previous sections appears here --- the grey line indicates the power-law $P^{1h} \propto b_g^{5}$.
In order to check whether this scaling is sensitive to the precise form of the HOD  
used, we also test a different HOD prescription for which: 
\be  
\bar{N}_g  \propto \log \frac{M}{M_{c}} \; \mathrm{for} \; M > M_c.
\label{Eq:AltHOD_1}
\ee 
The number of galaxies in each halo is drawn from a Poisson distribution with mean 
given by the formula above. The scaling obtained in this case between the one-halo 
term and the galaxy bias is shown on the bottom panel of Fig.~\ref{Fig:Bias_HOD_simu}. 
The grey line now follows $b_g^{4.5}$, a scaling very similar to the one found in the previous cases.

As is the case in the original HOD of the previous sections, in this alternative HOD prescription
the richness is an increasing function of halo mass. We also tested HOD models where the 
richness becomes constant, or even decreases, with increasing halo mass, and in those cases
we do not recover the same type of scaling between bias and the one-halo term of the power 
spectrum. 
However, we do not consider this to be a limitation of our study, as we expect realistic HODs 
to display an average number of galaxies monotonically increasing with halo mass. 


\section{Discussion}

We have shown that, when the galaxy bias is high enough ($b_g \gtrsim 3$), 
the 1-halo term of the power spectrum grows faster as a function of bias
than the 2-halo term. 
We also showed that the 1-halo term of the trispectrum grows much faster than
the 4-halo term. We argue that the 1-halo terms of all the 
N-polyspectra scale faster than the N-halo terms of those polyspectra.

We interpret these results in the following way.
Galaxy bias provides an intuitive physical interpretation 
of the way in which the visible matter distribution is related to 
the underlying DM distribution. Given a Gaussian field
$\delta_G$, taken from a distribution whose variance (in Fourier space) 
is basically the matter power spectrum, the density contrast of a biased 
tracer on a grid of finite-volume cells can be approximated by a 
lognormal field, $\delta_g = \exp [b_g \delta_G 
- b_g^2 \sigma_G^2/2] -1$, where $b_g$ is the bias of the tracer, 
and $\sigma_G^2$ is the variance of the Gaussian field on the volume of 
the cell of the grid \citep{ColesJones}. As the bias increases, the number of
particles found in the density peaks grow very fast
\citep{BBKS} --- exponentially, in the lognormal 
model --- while most of the space is emptied.

In terms of the Halo Model, we can vary the HOD parameters in such a way that 
the mean number density is kept fixed while the bias increases.
For highly biased galaxies we expect to find more objects concentrated in fewer halos. 
This implies that, for very high values of the bias, the galaxy 2-point statistics can 
get a large contribution from the statistics of halos hosting two or more
galaxies. This argument extends to all the $N$-point statistics:
in the high-bias limit there will be many galaxies inside the same few halos, 
which means that all the 1-halo terms of the $N$-point statistics will
become increasingly important.

In Section \ref{Subsec:PS}, within the \citet{PS} formalism,
we used very simple formulas for the halo richness and for the scaling of the mass
variance (and peak height), to show that the 1-halo term of the 
power spectrum grows very fast with bias.
A more refined analytical calculation, done in Section 
\ref{Subsec:ST} using the \citet{ST} framework, shows basically 
the same scaling.
In Section \ref{Sec:semi-analytic} we employed the \citet{Tinker} 
mass function, realistic HODs, and an exact computation of the mass
variance in $\Lambda$CDM models, and confirmed that the 1-halo term
of the power spectrum scales at least as fast as $P^{1h} \sim b_g^{4-5}$. 
Similarly, we showed that the 1-halo term of the trispectrum scales 
as fast as $T^{1h} \sim b_g^{12-15}$.
Finally, in Section \ref{Sec:simu} we used a halo catalog derived from the
 DEUS simulation, and two different kinds of HODs, and again obtained the 
 same scaling laws.

The fact that simple analytical arguments show the correct scaling 
of the 1-halo terms with bias is a hint that these results should be
related to basic properties of Gaussian fields and the matter power
spectrum. In fact, tracers of different biases effectively probe
different scales in the power spectrum: higher/lower masses (and higher/lower 
biases) are related to larger/smaller scales.
Since the matter power spectrum has a power index that 
ranges between $n \simeq -1$ at large scales, to $n\simeq -2$ at 
small scales, the power index which is relevant for the halo masses typical of a certain
type of galaxy is also an indicator of the bias of that galaxy.
Conversely, bias also tells us how galaxies are distributed among the 
dark matter halos, and about the typical peak height which corresponds
to a galaxy with that bias. 
In particular, a higher bias implies that galaxies will be more concentrated on
fewer halos, enhancing the 1-halo terms.
Since all these properties can be traced
back to the slope of the power spectrum, it is not surprising to 
find that different types of galaxies present the same scaling of the 
1-halo term with bias.

Since the 1-halo terms are constant on large scales,
for cosmological surveys that cannot resolve the inner structure 
of halos, these terms enter effectively as additional sources 
of noise and covariance. Conversely, very accurate and complete 
surveys will be able to detect much better the amplitudes and 
scale dependences of the 1-halo terms when the bias is sufficiently 
high, implying better constraints on HOD parameters.

In particular, cosmological surveys targeting highly-biased objects 
could be severely impacted by the
effects of the statistics of counts of the halos hosting those objects.
Measurements of the power spectrum from these surveys should take 
into account not only the higher effective shot noise coming 
from the 1-halo term of the power spectrum, but also the 
additional contribution to the power spectrum covariance coming 
from the 1-halo term of the trispectrum. Similarly, 
measurements of the bispectrum which employ highly biased 
tracers should ensure that the relevant 1-halo terms are properly 
taken into account. 

Finally, although we worked at $z=0$, it would be interesting to find out
what happens at high redshifts.
On the one hand, the bias of a given population of tracers is typically increasing
as a function of redshift; but on the other hand, linear theory should become a 
better approximation, which means that the 1-halo term should be less important. 
Therefore, at higher redshifts the form of the scaling of the 1-halo term should 
depend sensitively not only on the amplitude of the power spectrum, but also
on the evolution of the HOD parameters.

The cosmic infrared background (CIB) is a good example: in that case (redshifts $z\sim 1 - 4$), 
the 1-halo term has a higher amplitude compared with shot noise, which helped
constrain the HOD of the sources the CIB \citep{Thacker}. 
A similar situation may arise as galaxy surveys aim at deeper redshifts and higher 
number densities with objects such as emission-line galaxies and quasars 
[e.g., JPAS \citep{Benitez:2014}; eBOSS \citep{eBOSS}; DESI \citep{DESI}; PFS \citep{PFS}],
HI intensity maps (with, e.g., SKA\footnote{http://www.skatelescope.org/}), 
etc.: the amplitude of the 1-halo terms should be kept in check 
in order to ensure the reliability of the forecasts from these surveys.
We plan to examine in future work how extrapolations of the existing HODs 
to higher redshifts impact the signal-to-noise levels and the cosmological constraints 
for some future galaxy surveys.



\vskip 0.5cm

\noindent {\it  Acknowledgements}

We would like to thank Marcello Musso and Ravi Sheth for useful discussions,
and Aur\'elie P\'enin for many insightful comments.
We acknowledge support from the DIM ACAV of the Region \^Ile-de-France for the storage 
of the DEUS data.
We would also like to thank FAPESP, CAPES, CNPq, and the University of 
S\~ao Paulo's {\em NAP LabCosmos} for financial support. IB acknowledges 
support from FAPESP through the grant 2013/21069-9, and
FL acknowledges support from FAPESP through grants 2011/11973 and 2013/19936-6.

\bibliographystyle{mn2e.bst}


\begin{thebibliography}{99}

\bibitem[\protect\citeauthoryear{Abbott et~al.}{2005}]{Abbott:2005bi}
Abbott, T. {\em et~al.}, 2005, preprint (astro-ph/0510346)

\bibitem[\protect\citeauthoryear{Abell et~al.}{2009}]{LSST:2009pq}
Abell, P., {\em et~al.}, 2009, preprint (arXiv:0912.0201)

\bibitem[\protect\citeauthoryear{Abramo et~al.}{2012}]{2011arXiv1108.2657A}
Abramo, L.~R., {\em et~al.}, 2012, MNRAS, 423, 335



\bibitem[\protect\citeauthoryear{Adelman-McCarthy et~al.}{2008b}]{adelman-mccarthy_sixth_2008}
{Adelman-McCarthy}, J.~K.,  {\em et~al.}, 2008b, ApJS, 175, 297

\bibitem[\protect\citeauthoryear{Alimi et~al.}{2010}]{Alimi2010}
Alimi, J.-M., F\"uzfa, A., Boucher, V, Rasera, Y., Courtin, J., Corasaniti, P.~S., 2010, MNRAS, 401, 775

\bibitem[\protect\citeauthoryear{Anderson et~al.}{2012}]{Andersonetal2012}
{Anderson}, L.,  {\em et~al.}, 2012, MNRAS, 427, 3435

\bibitem[\protect\citeauthoryear{Anderson et~al.}{2014}]{Andersonetal2014}
{Anderson}, L.,  {\em et~al.}, 2014, MNRAS, 439, 83


\bibitem[\protect\citeauthoryear{Bardeen et~al.}{1986}]{BBKS}
Bardeen, J.~M., Bond, J.~R., Kaiser, N., Szalay, A.~S., 1986, ApJ, 304, 15 




\bibitem[\protect\citeauthoryear{Ben\'{\i}tez et~al.}{2009}]{Benitez:2008fs}
{Ben{\'{\i}}tez}, N., {\em et~al.}, 2009, ApJ, 691, 241

\bibitem[\protect\citeauthoryear{Ben\'{\i}tez et~al.}{2014}]{Benitez:2014}
{Ben{\'{\i}}tez}, N., {\em et~al.}, 2014, preprint (arXiv:1403.5237)

\bibitem[\protect\citeauthoryear{Berlind \& Weinberg}{2002}]{Berlind2002}
Berlind, A.~A., Weinberg, D.~H., 2002, ApJ 575, 587




\bibitem[\protect\citeauthoryear{Blake et~al.}{2011}]{2011MNRAS.415.2876B}
{Blake}, C., {\em et~al.}, 2011, MNRAS, 415, 2876




\bibitem[\protect\citeauthoryear{Carron}{2011}]{Carron2011}
{Carron}, J., 2011, ApJ, 738, 86

\bibitem[\protect\citeauthoryear{Carron}{2012}]{Carron2012a}
{Carron}, J., 2012, Phys. Rev. Lett., 108, 7

\bibitem[\protect\citeauthoryear{Carron \& Neyrinck}{2012}]{Carron2012b}
{Carron}, J., {Neyrinck}, M.~C., 2012, ApJ, 750, 1

\bibitem[\protect\citeauthoryear{Carron}{2014}]{Carron2014}
{Carron}, J., 2014, preprint (arXiv:1406.6095)

\bibitem[\protect\citeauthoryear{Cen \& Safarzadeh}{2015}]{CS2015}
{Cen}, R., {Safarzadeh}, M., 2015, ApJL, 798, 38


\bibitem[\protect\citeauthoryear{Chatterjee et~al.}{2013}]{Chatterjee2013}
{Chatterjee}, S., {Nguyen}, M.~L., {Myers}, A.~D., {Zheng}, Z., 2013, ApJ 779, 147


\bibitem[\protect\citeauthoryear{Cole et~al.}{2005}]{cole_2df_2005}
{Cole}, S., {\em et~al.}, 2005, MNRAS, 362, 505

\bibitem[\protect\citeauthoryear{Coles \& Jones}{1990}]{ColesJones}
{Coles}, P., Jones, B., 1990, MNRAS, 248, 1

\bibitem[\protect\citeauthoryear{Cooray \& Sheth}{2002}]{CooraySheth}
{Cooray}, A., {Sheth}, R., 2002, Phys. Rept., 372, 1

\bibitem[\protect\citeauthoryear{Courtin et~al.}{2011}]{Courtin2011}
Courtin, J., Rasera, Y., Alimi, J.-M., Corasaniti, P.~S., Boucher, V., F\"uzfa, A., 2011, MNRAS, 410, 1911




\bibitem[\protect\citeauthoryear{Dalal et~al.}{2008}]{NGDalal}
Dalal, N., Dor\'{e}, O., Huterer, D., Shirokov, A., 2008, Phys. Rev., D77, 123514

\bibitem[\protect\citeauthoryear{Dawson et~al.}{2012}]{BOSS}
Dawson, K., {\it et al.}, 2012, AJ, 145, 10

\bibitem[\protect\citeauthoryear{Dawson et~al.}{2015}]{eBOSS}
Dawson, K., {\it et al.}, 2015, preprint (arXiv:1508.04473)


\bibitem[\protect\citeauthoryear{Ellis et~al.}{2012}]{PFS}
Ellis, R., Takada, M., {\it et al.}, The PFS Team, 2012, preprint (arXiv:1206.0737)

\bibitem[\protect\citeauthoryear{Fry \& Gazta\~naga}{1993}]{FryGaz}
Fry, J., Gazta\~naga, E., 1993, ApJ, 413, 447


\bibitem[\protect\citeauthoryear{Kauffmann, White \& Guideroni}{1993}]{Kauffmann1993}
Kauffmann, G., White, S.~D.~M., Guiderdoni, B., 1993, MNRAS, 264, 201


\bibitem[\protect\citeauthoryear{Kauffmann et~al.}{1999}]{Kauffmann1999}
{Kauffmann}, G., {Colberg}, J.~M., {Diaferio}, A., {White}, S.~D.~M.
1999, MNRAS, 303, 188

\bibitem[\protect\citeauthoryear{Kayo \& Oguri}{2012}]{KO2012}
{Kayo}, I., {Oguri}, M., 2012, MNRAS, 424, 1363

\bibitem[\protect\citeauthoryear{Komatsu et~al.}{2009}]{WMAP5}
Komatso, E., et~al., 2009, ApJS, 180, 330



\bibitem[\protect\citeauthoryear{Kravtsov et~al.}{2004}]{Kravstov2004}
Kravtsov, A.~V., Berlind, A.~A., Wechsler, R.~H., Klypin, A.~A., Gottl\"ber, S., Allgood, B., Primack, J.~R., 2004, ApJ, 609, 35





\bibitem[\protect\citeauthoryear{Levi et~al.}{2013}]{DESI}
Levi, M., et~al., 2013, preprint (arXiv:1308.0847)

\bibitem[\protect\citeauthoryear{Linder}{2005}]{LinderGrowth}
Linder, E., 2005, Phys. Rev., D72, 043529

\bibitem[\protect\citeauthoryear{Ma \& Fry}{2000}]{MaFry2000}
Ma, C.-P., Fry, J.~N., 2000, ApJ, 543, 503

\bibitem[\protect\citeauthoryear{Maldacena}{2003}]{Maldacena2003}
Maldacena, J., 2003, JHEP, 5

\bibitem[\protect\citeauthoryear{Martinez \& Saar}{2001}]{MartinezBook}
Martinez, V., Saar, E., 2001, ``Statistics of the Galaxy Distribution'' 
(CRC Press, 2001)

\bibitem[\protect\citeauthoryear{Mo \& White}{1996}]{MoWhite1996}
{Mo}, H., {White}, S.~D.~M., 1995, MNRAS, 282, 347


\bibitem[\protect\citeauthoryear{Navarro, Frenk \& White}{1995}]{Navarro1995}
{Navarro}, J.~F. and {Frenk}, C.~S., {White}, S.~D.~M., 1995, MNRAS, 275, 56


\bibitem[\protect\citeauthoryear{Navarro, Frenk \& White}{1997}]{NFW}
{Navarro}, J.~F. and {Frenk}, C.~S. and {White}, S.~D.~M., 1997, ApJ, 490, 483


\bibitem[\protect\citeauthoryear{PAN-STARRS}{}]{PAN-STARRS}
Tonry, J.~L., Stubbs, C.~W., Lykke, K.~R., et al., 2012, ApJ, 750, 99

\bibitem[\protect\citeauthoryear{Planck Collaboration}{2015}]{Planck2015-cosmo}
Planck Collaboration, 2015, preprint (arXiv:1502.01589)

\bibitem[\protect\citeauthoryear{Piscionere et~al.}{2014}]{Piscionere2014}
Piscionere, J.~A., Berlind, A.~A., McBride, C.~K., Scoccimarro, R., 2014, 1407.6740

\bibitem[\protect\citeauthoryear{Porciani, Magliocchetti, \& Norberg}{2004}]{Porciani}
Porciani, C., Magliocchetti, M., Norberg, P. 2004, MNRAS, 355, 1010

\bibitem[\protect\citeauthoryear{Press \& Schechter}{1974}]{PS}
{{Press}, W.~H., {Schechter}, P.}, 1974, ApJ, 187, 425
    
\bibitem[\protect\citeauthoryear{Rasera et~al.}{2010}]{Rasera2010} Rasera, Y., Alimi, J.-M., Courtin, J., Roy, F., Corasaniti, P.-S., F\"uzfa, A., Boucher, V., 2010, INVISIBLE UNIVERSE: Proceedings of the Conference. AIP Conference Proceedings, 1241, 1134-1139

\bibitem[\protect\citeauthoryear{Richardson et~al.}{2012}]{Richardson2012}
{Richardson}, J.,  Zheng, Z., Chatterjee, S., Nagai, D., Shen, Y.,
2012, ApJ, 755, 30



\bibitem[\protect\citeauthoryear{Schlegel et~al.}{2009}]{BigBOSS}
Schlegel, D.~J., et~al., 2009, preprint (arXiv:0904.0468)

\bibitem[\protect\citeauthoryear{Scoville et~al.}{2007}]{scoville_cosmic_2007}
Scoville, N., {\em et~al.}, 2007, ApJS, 172, 1


\bibitem[\protect\citeauthoryear{Seljak}{2000}]{SeljakHOD}
Seljak, U., 2000, MNRAS, 318, 203


\bibitem[\protect\citeauthoryear{Shen et~al.}{2007}]{Shen2007}
Shen, Y., et~al., 2007, AJ, 133, 2222

\bibitem[\protect\citeauthoryear{Shen et~al.}{2010}]{Shen2010}
Shen, Y., et~al., 2010, ApJ, 719, 2

\bibitem[\protect\citeauthoryear{Shen et~al.}{2013}]{Shen2013}
Shen, Y., et~al., 2013, ApJ, 778, 98

\bibitem[\protect\citeauthoryear{Sheth, Mo \& Tormen}{1999}]{SMT}
{{Sheth}, R.~K.,{Mo}, H.~J., {Tormen}, G.}, 2001, MNRAS, 323, 1


\bibitem[\protect\citeauthoryear{Sheth \& Tormen}{1999}]{ST}
{{Sheth}, R.~K., {Tormen}, G.}, 1999, MNRAS, 308, 119
    



\bibitem[\protect\citeauthoryear{Springel et~al.}{2005}]{Springel2005}
Springel, V., et~al., 2005, Nature, 435, 629




\bibitem[\protect\citeauthoryear{Thacker et~al.}{2013}]{Thacker}
Thacker, C., et al., 2013, ApJ, 768, 58

\bibitem[\protect\citeauthoryear{Tinker et~al.}{2008}]{Tinker}
{{Tinker}, J., {Kravtsov}, A.~V., {Klypin}, A., {Abazajian}, K.,
{Warren}, M., {Yepes}, G., {Gottl{\"o}ber}, S., {Holz}, D.~E.},
2008, ApJ, 688, 709

\bibitem[\protect\citeauthoryear{Tinker et~al.}{2010}]{Tinker10}
{{Tinker}, J.~L. and {Robertson}, B.~E. and {Kravtsov}, A.~V. and 
{Klypin}, A. and {Warren}, M.~S. and {Yepes}, G. and {Gottl{\"o}ber}, S.},
2010, ApJ, 724, 878




\bibitem[\protect\citeauthoryear{Wake et~al.}{2008}]{Wake2008}
{Wake}, D.~A., {Croom}, S.~M., {Sadler}, E.~M., {Johnston}, H.~M., 2008, MNRAS, 391, 1674


\bibitem[\protect\citeauthoryear{Watson et~al.}{2010}]{Watson2010}
Watson, D.~F., Berlind, A.~A., McBride, C.~K., Masjedi, M., 2010, ApJ, 709, 115

\bibitem[\protect\citeauthoryear{Weinberg}{2002}]{BiasWeinberg}
Berlind, A.~A., Weinberg, D.~H., 2002, ApJ, 575, 587 

\bibitem[\protect\citeauthoryear{White \& Frenk}{1991}]{WF1991}
White, S., Frenk, C., 1991, ApJ, 379, 52 


\bibitem[\protect\citeauthoryear{York et~al.}{2000}]{York:2000gk}
York, D.~G., {\em et~al.}, 2000, AJ, 120, 1579

\bibitem[\protect\citeauthoryear{Zentner, Hearin \& van den Bosch}{2014}]{Zentner2014}
Zentner, A. R., Hearin, A. P., van den Bosch, F. C., 2014,
MNRAS, 443, 3044


\bibitem[\protect\citeauthoryear{Zheng et~al.}{2005}]{Zheng2005}
Zheng, Z., {\em et~al.}, 2005, ApJ, 633, 791

\bibitem[\protect\citeauthoryear{Zheng, Coil \& Zehavi}{2007}]{Zheng2007}
Zheng, Z.,  Coil, A., Zehavi, I., 2007, ApJ, 667, 760

\bibitem[\protect\citeauthoryear{Zheng et~al.}{2009}]{Zheng2009}
Zheng, Z., Zehavi, I., Eisenstein, D. J., Weinberg, D. H., Jing,
Y. P., 2009, ApJ, 707, 554

\end{thebibliography}




\end{document}